\documentclass[preprint,3p,times]{elsarticle}

\usepackage{amsfonts,amsmath,amsthm,amssymb}
\usepackage{graphicx,epsfig,pstricks,pst-node}
\usepackage{lineno}
\usepackage{figsize}
\usepackage{stfloats}
\usepackage{color,xcolor}
\usepackage{url, hyperref}

\journal{12345}

\begin{document}

\begin{frontmatter}

\title{Edge-based reputation promotes cooperation in simplicial complexes}
\author[inst1]{Chunpeng Du}
\author[inst1]{Fei Fang}
\author[inst2]{Alfonso de Miguel-Arribas}
\author[inst3]{Yikang Lu}
\ead{luyikang\_top@163.com}
\author[inst4]{Yanan Wang}
\ead{yanan.wang@buaa.edu.cn}
\author[inst4]{Xin Pan}
\author[inst5,inst6]{Yamir Moreno}
\ead{yamir@unizar.es}

\affiliation[inst1]{organization={School of Mathematics},
            addressline={Kunming University}, 
            city={Kunming},
            postcode={650214}, 
            state={Yunnan},
            country={China}}   

\affiliation[inst2]{organization={Zaragoza Logistics Center (ZLC)},
            city={Zaragoza},
            postcode={50018}, 
            state={Zaragoza},
            country={Spain}}  

\affiliation[inst3]{organization={School of Statistics and Mathematics},
addressline={Yunnan University of Finance and Economics}, 
            city={ Kunming},
            postcode={650221}, 
            state={Yunnan},
            country={China}}  

\affiliation[inst4]{
	organization={School of Reliability and Systems Engineering},
	addressline={Beihang University},             
	city={Beijing},            
	postcode={100191},             
	state={Beijing},            
	country={China}}

\affiliation[inst5]{
	organization={SInstitute for Biocomputation and Physics of Complex Systems (BIFI)},
	addressline={University of Zaragoza},             
	city={Zaragoza},            
	postcode={50018},                       
	country={Spain}}

\affiliation[inst6]{
	organization={Department of Theoretical Physics},
	addressline={University of Zaragoza},             
	city={Zaragoza},            
	postcode={50018},                       
	country={Spain}}

\begin{abstract}
Understanding how cooperation emerges and persists is a central challenge in the evolutionary dynamics of social and biological systems. Most prior studies have examined cooperation through pairwise interactions, yet real-world interactions often involve groups and higher-order structures. Reputation is a key mechanism for guiding strategic behavior in such contexts, but its role in higher-order networks remains underexplored. In this study, we introduce an edge-based reputation mechanism, incorporating both direct and indirect reputation, to investigate the evolution of cooperation in simplicial complexes. Our results show that coupling reputation mechanisms with higher-order network structures strongly promotes cooperation, with direct reputation exerting a stronger influence than indirect reputation. Moreover, we reveal a nonlinear interplay between network topology and reputation mechanisms, highlighting how multi-level structures shape collective outcomes. These findings provide a novel theoretical framework for understanding cooperation in complex social systems.
\end{abstract}


\end{frontmatter}

\section{Introduction}
\label{sec:intro}

Cooperative behavior, characterized by actions where individuals incur costs to benefit others, is a pervasive phenomenon observed across social species, from microorganisms to humans~\cite{nowak2012evolving,xia2011enhancement,perc2017statistical,santos2012role,chen2015competition}. Darwin’s theory of natural selection, however, poses challenges in explaining the emergence and persistence of such behavior~\cite{perc2016phase,nowak2004emergence}. To elucidate how cooperation evolves under selective pressures, researchers have proposed diverse mechanisms underpinning its evolutionary dynamics~\cite{smith1964group,lieberman2005evolutionary,trivers1971evolution,axelrod1981emergence}. In 2006, Nowak synthesized these efforts, delineating five fundamental mechanisms that promote cooperation: kin selection, group selection, direct reciprocity, indirect reciprocity, and network reciprocity~\cite{nowak2006five}.

In the study of cooperation evolution, network reciprocity has emerged as a key theoretical framework for understanding how local interactions drive global cooperative dynamics~\cite{chen2014optimal,perc2013evolutionary,xia2014evolution,chen2008promotion}, owing to its ability to effectively capture the spatial distribution of cooperative behavior in structured populations~\cite{wang2013interdependent,xia2015heterogeneous,chen2018punishment}. However, traditional network reciprocity models, which are confined to pairwise interactions, face significant limitations in explaining collective behaviors, particularly their failure to account for synergistic effects in groups of three or more individuals~\cite{perc2006double,alvarez2021evolutionary}. This has spurred the development of higher-order network frameworks, which provide a more general and comprehensive approach~\cite{sheng2024strategy,iacopini2024temporal,civilini2024explosive}. By employing mathematical structures such as simplicial complexes, these higher-order networks directly model group interactions, enabling precise characterization of nonlinear effects arising from synergy and offering a robust representation of complex reciprocal relationships among multiple agents~\cite{majhi2022dynamics,zhang2023higher,santos2006evolutionary,pacheco2009evolutionary}. As a result, this framework establishes a realistic topological foundation for analyzing information propagation and strategy evolution within groups. Building on these theoretical advantages, recent studies have systematically explored the roles of factors such as the prevalence of higher-order interactions~\cite{guo2021evolutionary}, embedded social norms~\cite{ma2024social}, and network heterogeneity~\cite{wang2024mixing} in shaping cooperation dynamics within this paradigm.

The reputation mechanism, a critical driver of cooperation evolution~\cite{xia2023reputation,guo2018reputation,xia2020effect,chen2014solving,wang2012inferring,xia2016risk}, remains largely underexplored within higher-order network frameworks. While its synergistic role in traditional networks is well-documented (in regular networks, reputation enhances local interactions to facilitate cooperator clustering~\cite{gong2020reputation,dong2019second}; in Erdős–Rényi networks, highly connected topologies accelerate reputation dissemination across populations~\cite{podder2021local,chen2007prisoner}), these insights are limited to simplistic interaction models and have not been adequately extended to higher-order structures. This gap arises partly from the structural constraints of conventional network models, partly from the reliance of traditional reputation mechanisms on oversimplified evaluation schemes~\cite{murase2022social}. These schemes include binary classifications (e.g., cooperation versus defection), basic scoring systems (e.g., `good' or `bad' reputation), and weighted aggregations of past behaviors (e.g., tit-for-tat in direct reciprocity)~\cite{perc2012sustainable,wang2013reputation,chen2018prisoner}. Although effective in specific scenarios, such approaches fail to capture the intricate, dynamic interactions individuals encounter across multiple groups in complex social networks. Consequently, within higher-order network frameworks, developing innovative reputation mechanisms and assessing their influence on cooperation evolution has become an urgent and compelling research priority.

Given the limitations of traditional network models and reputation mechanisms, we introduce a novel edge-based reputation framework for higher-order networks. Unlike conventional node-based approaches, it assigns each edge two independent reputation vectors. In the interaction between nodes $i$ and $j$, $R_{\vec{ij}}$ measures $i$'s assessment of $j$ based on $j$'s past actions toward $i$, while $R_{\vec{ji}}$ captures the reverse. This bidirectional design offers a finer-grained view of interaction dynamics than single-value reputation models. Building on Bandura's social cognitive theory and the Nobel Prize nomination process \cite{bandura2001social}, we distinguish between direct and indirect reputation. Direct reputation arises from interactions with neighbors and reflects subjective evaluations. Indirect reputation stems from shared neighbors, aggregating their views into an objective assessment. This dual mechanism captures trust propagation in complex social systems and advances the application of reputation in higher-order network models within evolutionary game theory.

The edge reputation mechanism proposed in this study integrates direct and indirect reputation, significantly enhancing the precision and dimensionality of reputation evaluation compared to traditional single-reputation mechanisms. Through evolutionary game theory and numerical simulations, we systematically investigate the role of the reputation mechanism across diverse network topologies and multiple social dilemmas. Our findings reveal that indirect reputation sustains cooperation by integrating information, while direct reputation markedly promotes cooperation due to its real-time updating characteristics. Furthermore, the study uncovers a nonlinear coupling effect between higher-order network structures and the reputation mechanism. These findings provide the first insight into the propagation dynamics of reputation mechanisms in higher-order networks, offering a novel theoretical framework for understanding the emergence of cooperative behavior in complex social networks.

This paper is organized as follows: In Sec.~\ref{sec:model}, we introduce the edge reputation model within the framework of higher-order networks. Then, in Sec.~\ref{sec:results}, we systematically investigate the role of the edge reputation mechanism in higher-order networks and present the associated findings. Finally, in Sec.~\ref{sec:conclusion}, we summarize the findings and discuss their further implications.

\section{Model}
\label{sec:model}

Guo et al.~\cite{guo2021evolutionary} extended evolutionary games from pairwise networks to simplicial complexes. In this work, we augment this framework by studying how edge-based reputations shape cooperation. For clarity, we recapitulate in some detail how games are defined in these higher-order settings, and make our contribution clear when introduced.

\begin{figure}
    \centering
    \includegraphics[width=0.7\linewidth]{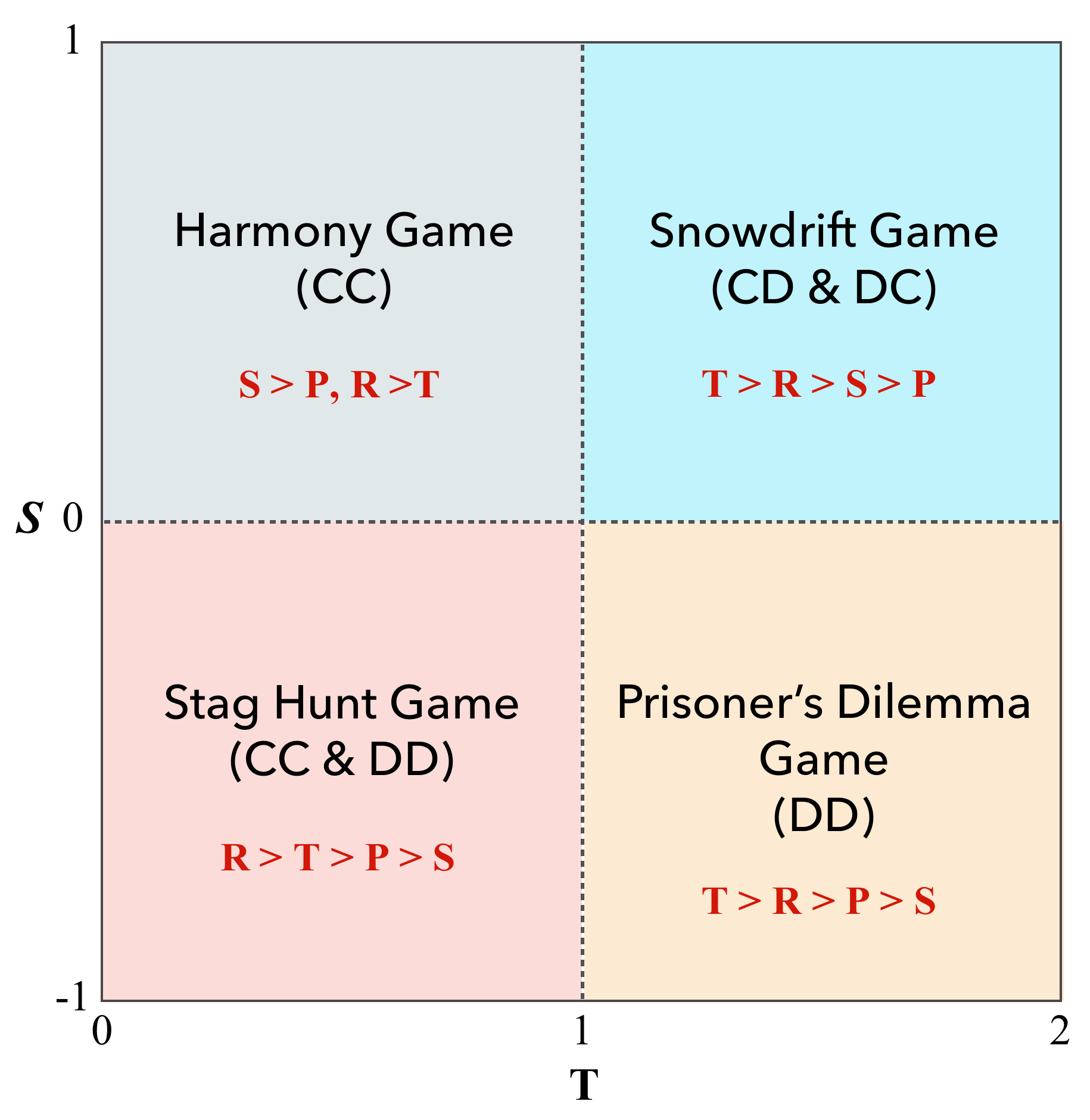}
   \caption{\textbf{Social dilemma games diagram.} Schematic representation of the games with free parameters $(T, S)$. The games are categorized into four quadrants based on the conditions $0 \leq T \leq 2$ and $-1 \leq S \leq 1$. From left to right and top to bottom: the Harmony game, the Snowdrift game, the Stag Hunt game, and the Prisoner's Dilemma game. For illustrative purposes, the reward (R) and punishment (P) values have been fixed to be $1$ and $0$, respectively. The annotations in parentheses represent the strategies corresponding to the Nash equilibria of the $2$-player games, where $C$ denotes cooperation and $D$ denotes defection. We also show the ordering of the entries $(R,S,T,P)$ of the payoff matrix for each game.}
    \label{fig:1}
\end{figure}

We assume that individuals play a two-strategy game, with strategies comprising \textit{cooperation} ($C$ ) and \textit{defection} ($D$), with payoffs $(R,S,T,P)$ specifying the underlying social dilemmas: the Harmony (HG), the Snowdrift (SD), the Stag Hunt (SH), and the Prisoner's Dilemma (PD) games. In pairwise encounters, cooperation yields $R$ to both players, mutual defection yields $P$, and in mixed pairs the cooperator receives $S$ while the defector receives $T$. The ordering of $(R,S,T,P)$ determines which social dilemma is in effect and the associated Nash equilibria. Figure \ref{fig:1} depicts the games' positioning in a $(T,S)$ diagram, for fixed $R=1$ and $P=0$, as well as the ordering of the entries of the payoff matrix for each of the dilemmas.

\medskip
\noindent
\textbf{Population structure and strategies.}
Nodes represent players, and the topology of connections that give rise to the population structure is a simplicial complex whose $1$-skeleton (the underlying graph) is the contact network. An edge $\{i,j\}$ is a $1$-simplex, and a filled triangle $\{i,j,k\}$ is a $2$-simplex~\cite{battiston2020networks}. We distinguish $3$-cycles (open triangles in the $1$-skeleton, with no attached $2$-simplex) from $2$-simplices (filled triangles encoding a higher-order, triadic interaction). A structural parameter $\rho\in[0,1]$ controls the proportion of higher-order interactions in the system: $\rho=0$ yields a purely pairwise structure (only $1$-simplices), while $\rho=1$ assigns a $2$-simplex to every $3$-cycle. The specific details of the network structure and generative model can be found in \cite{guo2021evolutionary, kovalenko2021growing}.

\begin{figure}[htbp]
    \centering
    \includegraphics[width=0.9\linewidth]{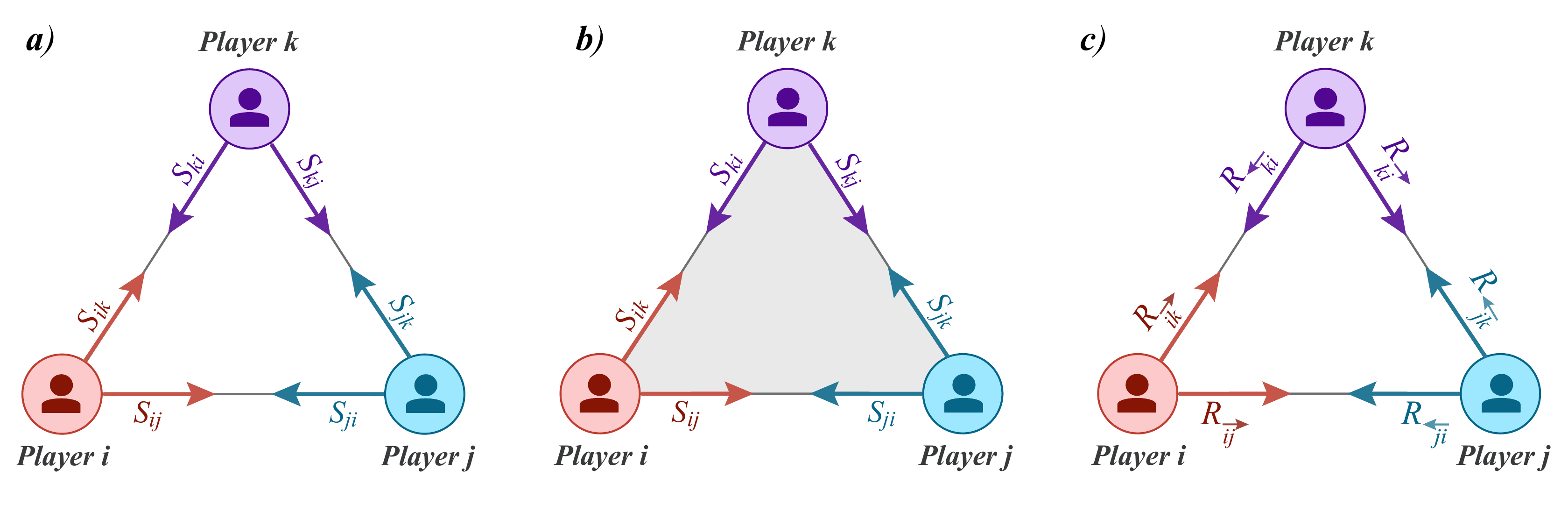}
    \caption{\textbf{Interaction topology schematics.} Dyadic closure (a), triadic interactions (b), and edge reputation among nodes-players $i$, $j$, and $k$ (c). Dyadic closure results in a triangle devoid of higher-order properties (1-simplex), whereas triadic interactions give rise to a filled triangle (2-simplex). The entry $s_{ij}$ denotes the strategy adopted by node $i$ toward node $j$. The strategy set is binary, comprising $0$ ($\equiv C$, for cooperation) and $1$ ($\equiv D$, for defection). Edge reputation (c) is represented by $R_{\vec{ij}}$, node $i$’s assessment of node $j$, and $R_{\vec{ji}}$, node $j$’s assessment of node $i$.}
    \label{fig:2}
\end{figure}

A strategy matrix $\mathbf{S}=\{s_{ij}\}$ is introduced to describe interactions between players $i$ and $j$. Here, $s_{ij}=0$ signifies cooperation $C$ by player $i$ toward player $j$, while $s_{ij}=1$ denotes defection $D$. Each player $i$ may adopt distinct strategies toward different neighbors, forming a strategy vector with dimension equal to their degree $k_i$. The degree $k_i$ is expressed as $k_i=k_i^C+k_i^D$, where $k_i^C$ represents the number of neighbors with whom player $i$ cooperates, and $k_i^D$ indicates the number of neighbors against whom player $i$ defects. Similarly, each player $i$ assigns a reputation assessment to their neighbors based on historical behavior. For instance, $R_{\vec{ij}}$ quantifies node $i$'s reputation assessment of node $j$, shaped by $j$'s past actions toward $i$. Fig.~\ref{fig:2}(a) and Fig.~\ref{fig:2}(b) illustrate, respectively, the case of a dyadic closure and that of a triadic interaction, while Fig.~\ref{fig:2}(c) depicts edge reputations between nodes.

\medskip
\noindent
\textbf{Payoff structure}. Given the higher-order setting, the payoff computation is more convoluted than in classical pairwise interactions. While players are endowed with a total payoff $\Pi_i$ per round, its calculation is not based on accumulation through pairwise interactions with their neighbors, but through edge-based interactions in the simplicial structure. Thus, this is performed as
\begin{equation}
\Pi_i = \frac{1}{k_i}\sum_{j\in \Omega(i)} \Pi_{i,(ij)},
\label{eq:cumulative_payoff}
\end{equation}
where $\Omega(i)$ represents the set of neighbors of node $i$, $\Pi_{i,(ij)}$. Here, $\Pi_{i,(ij)}$ is the cumulative payoff of node $i$ along the specific edge $(ij)$. In turn, this quantity is computed as accumulated through the payoff $ \Pi_{i,(ij),\tau}$ of each triangle $\tau$ involving the edge $(ij)$:
\begin{equation}
\Pi_{i,(ij)}=\frac{1}{k_{ij}}\sum_{\tau \in \Delta} \Pi_{i,(ij),\tau},
\label{eq:edge_payoff}
\end{equation}
where $\Delta$ denotes the set of all $k_{ij}$ triangles involving the edge $(ij)$, and $\Pi_{i,(ij),\tau}$ is the payoff of player $i$ along the specific edge $(ij)$ with respect to the specific triangle $\tau$. Now, in a classical pairwise setting, payoffs would be obtained from the $2\times 2$ payoff matrix such as:
 \begin{equation}
        \begin{array}{c@{\hspace{0pt}}c}& 
        \begin{array}{cc}
            C & D
        \end{array}
        \\
        \begin{array}{c}
        C \\
        D
        \end{array}
        &
        \begin{pmatrix}
        1 & S \\
        T & 0
        \end{pmatrix}
        \end{array}
        \label{eq:payoff_matrix}
    \end{equation}
However, as recognized in our guiding reference \cite{guo2021evolutionary}, it is crucial to distinguish between the case in which a given triangle $\tau$ represents just a dyadic closure (\ref{fig:2}a) and the case of a $2$-simplex (\ref{fig:2}b). This consideration gives rise to $3$ types of possible games to be played based on the type of triangle $\tau$ to which a specific interaction edge $(ij)$ belongs to:
\begin{itemize}
    \item \textbf{Open triangle}: Triangle $\tau$
    is open, i.e., a closure of three $1$-simplices, then the edge $(ij)$ represents a pairwise interaction between players $i$ and $j$. The payoff $\Pi_{i,(ij),\tau}$ is given by Eq.~\ref{eq:payoff_matrix}, with $S=S_1$ and $T=T_1$. This sets \textbf{Game~1}.
    \item \textbf{Triadic interaction}: Triangle $\tau$ results from a triadic interaction, meaning that edge $(ij)$ is part of a $2$-simplex. Subsequently, the obtention of $\Pi_{i,(ij),\tau}$ needs to explicitly involve the strategy of the remaining player in the $2$-simplex, $k$, implying the introduction of a tensor. Then, we have to check the strategies $s_{ki}$ and $s_{kj}$ of node $k$ towards neighbors $i$ and $j$. Then:
    \begin{itemize}
        \item If $s_{ki}=s_{kj}$ (aligned), then players $i$ and $j$ play \textbf{Game~2} along edge $(ij)$. Payoffs are computed with Eq.~\ref{eq:payoff_matrix} with $S=S_2$ and $T=T_2$. 
        \item If $s_{ki}\neq s_{kj}$ (misaligned), players  $i$ and $j$ play \textbf{Game~3} along edge $(ij)$. Now, payoffs are computed with Eq.~\ref{eq:payoff_matrix} with $S=S_3$ and $T=T_3$. 
    \end{itemize}   
\end{itemize}

Intuitively, in a $2$-simplex the third player's stance toward $i$ and $j$ modulates how the pair perceives their interaction: aligned third-party behavior induces one payoff setting, misaligned behavior another. The total payoff on $(i,j)$ is the average over its incident triangles, and $\Pi_i$ is the average over all $(i,j)$ with $j\in N_i$.

\medskip
\noindent
\textbf{Reputation dynamics.}
Here, we present our main contribution in the modeling of evolutionary games on simplicial complexes by introducing both direct and indirect reputation mechanisms.

The directed reputation value $R_{\vec{ij}}(t)$ represents the evaluation of player $j$ by player $i$ at iteration $t$. It updates according to player $j$'s strategy toward $i$ in the previous step and the number of triangles adjacent to edge $(i,j)$. Cooperation by $j$ increases $R_{\vec{ij}}$ by $\Delta R_{\vec{ij}}(t-1)$; defection decreases it by the same amount, where
\begin{equation}
\Delta R_{\vec{ij}}(t-1) = \Delta \mathrm{Tr}^{\mathrm{one}}_{ij} + \Delta \mathrm{Tr}^{\mathrm{two}}_{ij}.
\label{eq:reputation_delta}
\end{equation}
In this context, the triangles adjacent to edge $(i,j)$ comprise both open 3-cycles (1-simplices) and filled $2$-simplices.  
Accordingingly, $\Delta \mathrm{Tr}^{\mathrm{one}}_{ij}$ collects the contribution from interactions where $(i,j)$ participates in open triangles (1-simplices, i.e., Game~1),  
whereas $\Delta \mathrm{Tr}^{\mathrm{two}}_{ij}$ aggregates the contribution from interactions where $(i,j)$ belongs to filled triangles (2-simplices, i.e., triadic interactions in Games~2 and~3). 
Thus, $\Delta R_{\vec{ij}}$ is the total local reputation increment accumulated by $j$ along edge $(i,j)$ over all such adjacent triangles.
A decay factor $\alpha$ reflects the fading influence of past actions:
\begin{equation}
R_{\vec{ij}}(t) =  
\begin{cases}  
\alpha R_{\vec{ij}}(t-1) + \Delta R_{\vec{ij}}(t-1), & \text{if } s_{ji}=0,\\[4pt]  
\alpha R_{\vec{ij}}(t-1) - \Delta R_{\vec{ij}}(t-1), & \text{if } s_{ji}=1.
\end{cases}  
\label{eq:reputation_update}
\end{equation}

\begin{figure}[htbp]
    \centering
    \includegraphics[width=0.7\linewidth]{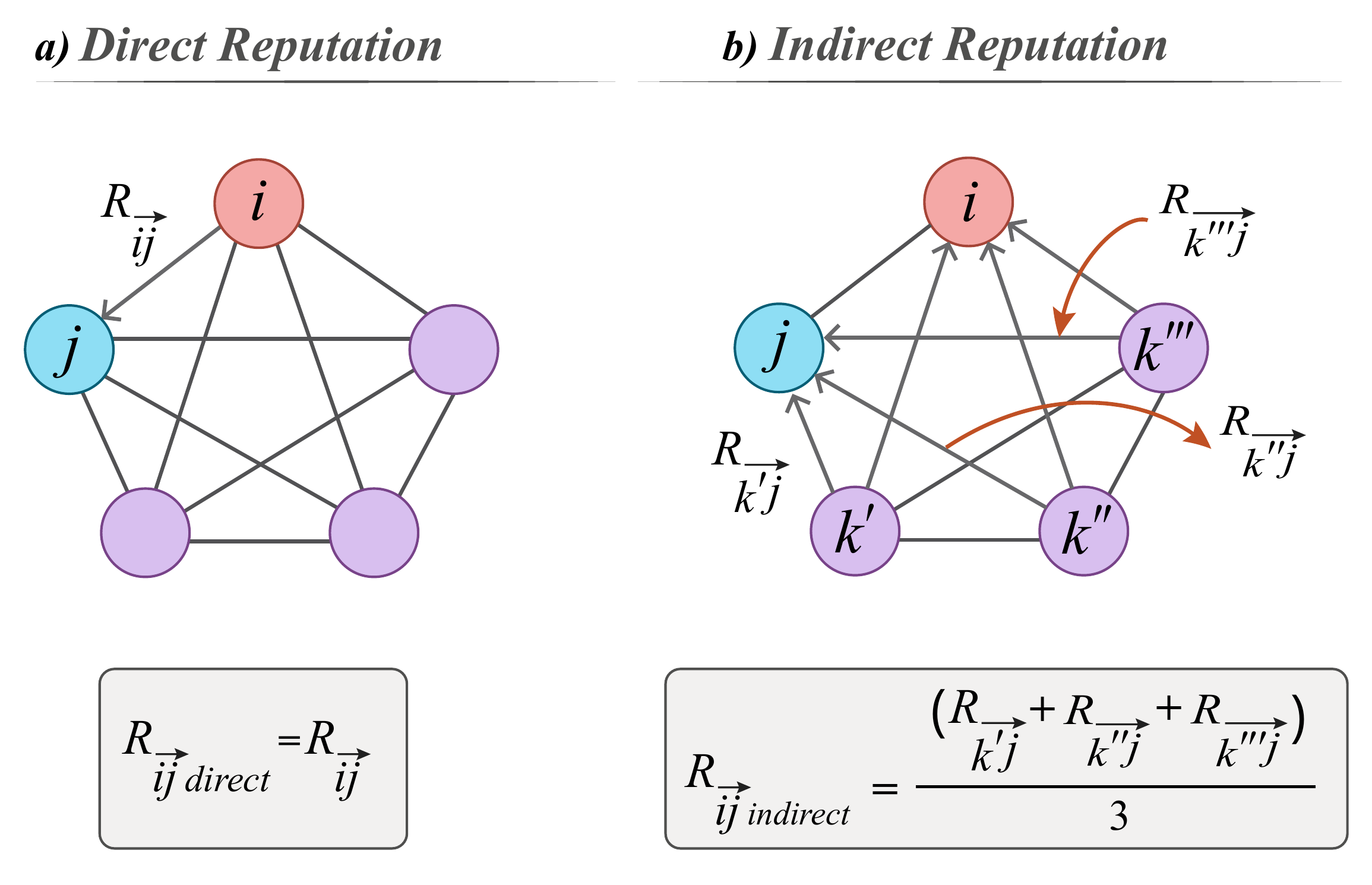}
    \caption{\textbf{Reputation assessment mechanism.} 
    Left: Player $i$'s direct reputation assessment of neighbor $j$, $R^{\mathrm{direct}}_{\vec{ij}}$, derived from $j$'s historical behavior toward $i$. 
    Right: Indirect reputation assessment, $R^{\mathrm{indirect}}_{\vec{ij}}$, obtained by averaging assessments of $j$ provided by common neighbors $k'$, $k''$, and $k'''$.}
    \label{fig:3}
\end{figure}

Each player $i$ evaluates every neighbor $j$ by integrating direct and indirect assessments:
\begin{equation}
R^{\mathrm{eval}}_{\vec{ij}}(t)
= \gamma\, R^{\mathrm{direct}}_{\vec{ij}}(t)
+ (1-\gamma)\, R^{\mathrm{indirect}}_{\vec{ij}}(t),
\qquad \gamma \in [0,1].
\label{eq:reputation_assessment}
\end{equation}
The components are defined as
\begin{align}
R^{\mathrm{direct}}_{\vec{ij}}(t) &= R_{\vec{ij}}(t), \label{eq:direct_term}\\[4pt]
R^{\mathrm{indirect}}_{\vec{ij}}(t) &= 
\frac{1}{|N_i \cap N_j|} 
\sum_{k \in N_i \cap N_j} R_{\vec{kj}}(t),
\label{eq:indirect_term}
\end{align}
for $N_i \cap N_j \neq \emptyset$, where $N_i \cap N_j$ denotes the set of common neighbors shared by players $i$ and $j$, and $R_{\vec{kj}}$ indicates the assessment of $j$ by a common neighbor $k$.

\medskip
\noindent
\textbf{Direct versus indirect reputation.} 
Although both reputation components are updated synchronously at each simulation step and each retains memory through the decay factor~$\alpha$, their informational bases differ fundamentally. 
\textbf{Direct reputation} $R^{\mathrm{direct}}_{\vec{ij}}(t)$ summarizes player~$i$'s own historical interactions with~$j$, providing a first-hand and highly responsive evaluation. 
\textbf{Indirect reputation} $R^{\mathrm{indirect}}_{\vec{ij}}(t)$, in contrast, aggregates the reputational assessments of~$j$ made by their common neighbors $k \in N_i \cap N_j$. 
Each of these neighboring assessments $R_{\vec{kj}}(t)$ already embeds $k$'s experience with~$j$, so the information reaching~$i$ through this pathway is effectively second-hand and filtered through the network. 
This additional layer of mediation produces a slower, more diffused adaptation to behavioral changes—an \textit{effective informational lag} rather than an explicit temporal delay.

\medskip
\noindent
Using these evaluations, each player $i$ selects the neighbor with the highest total reputation assessment, denoted $j_{\max}$, as the exemplar for learning their strategy. 
The whole set of strategies $s_{ij}$, $j=1,...,k_i$, of player $i$ are then updated according to the Fermi rule:
\begin{equation}
W = \frac{1}{1 + \exp\!\left(\frac{\Pi_i - \Pi_{j_{\max}}}{K}\right)}.
\label{eq:fermi_update}
\end{equation}
Thus, in case of successful updating, player $i$ simultaneously updates all their $k_i$ strategies.

\medskip
\noindent
\textbf{Monte Carlo simulations.} 
To investigate the evolutionary dynamics of cooperation under the proposed framework, we perform extensive Monte Carlo simulations on simplicial complexes generated through a preferential attachment mechanism, following Guo et al.~\cite{guo2021evolutionary} and Kovalenko et al.~\cite{kovalenko2021growing}. Each node represents a player and each edge $(ij)$ carries both a strategy $s_{ij}$ and a reputation value $R_{\vec{ij}}$. Initially, all strategies and reputations are randomly assigned, with reputation values bounded within $[0,100]$ to ensure stability.
Importantly, these bounds for the reputation values are preserved during the simulations, and thus reputation values exceeding the range are capped at $100$, whereas values going negative are set to zero.


The system evolves over $2\times10^4$ Monte Carlo steps (MCS). During each MCS, every player $i$ sequentially:
\begin{enumerate}
    \item Plays the corresponding social dilemma game with each neighbor $j$, where payoffs depend on the type of triangle $\tau$ associated with edge $(i,j)$, as defined in Eqs.~\ref{eq:cumulative_payoff}--\ref{eq:edge_payoff}.
    \item Accumulates payoffs $\Pi_i$ according to Eq.~\ref{eq:cumulative_payoff}.
    \item Updates edge reputations $R_{\vec{ij}}$ based on neighbors' previous actions using Eq.~\ref{eq:reputation_update}, with decay factor $\alpha=0.9$.
    \item Evaluates each neighbor's reputation through Eq.~\ref{eq:reputation_assessment}, combining direct and indirect assessments.
    \item Selects the neighbor with the highest evaluated reputation, $j_{\mathrm{max}}$, as the learning model, and updates strategies toward all neighbors using the Fermi rule (Eq.~\ref{eq:fermi_update}), with noise parameter $K=0.01$.
\end{enumerate}

To prevent premature convergence, a small mutation probability allows spontaneous random strategy flips, following Shen et al.~\cite{shen2025mutation}. Unless otherwise stated, simulations use a population size of $N=500$, and the baseline payoff parameters $R=1$ and $P=0$. The system is iterated until a stationary regime is reached; the average cooperation level is then computed over the last $3\times10^3$ MCS to characterize steady-state behavior. All results are averaged over five independent network realizations.

\section{Results}
\label{sec:results}

Firstly, to facilitate the analysis of this model, we standardized the parameters of Game~1 and Game~3 (i.e., setting $T_1=T_3$ and $S_1=S_3$) while varying the parameters of Game~2 ($T_2$,$S_2$). Given the paradigmatic significance and inherent complexity of the PD in social dilemma research, our initial analysis focuses on the scenario where both Game~1 and Game~3 are configured as Prisoner's Dilemmas. By setting $S_2 = -0.5$ and $T_2 \in [0,2]$, Game 2 can be tuned to represent either a stag hunt ($T_2\in[0,1]$) or a prisoner's dilemma ($T_2\in[1,2]$).

\begin{figure}
    \centering
    \includegraphics[width=0.9\linewidth]{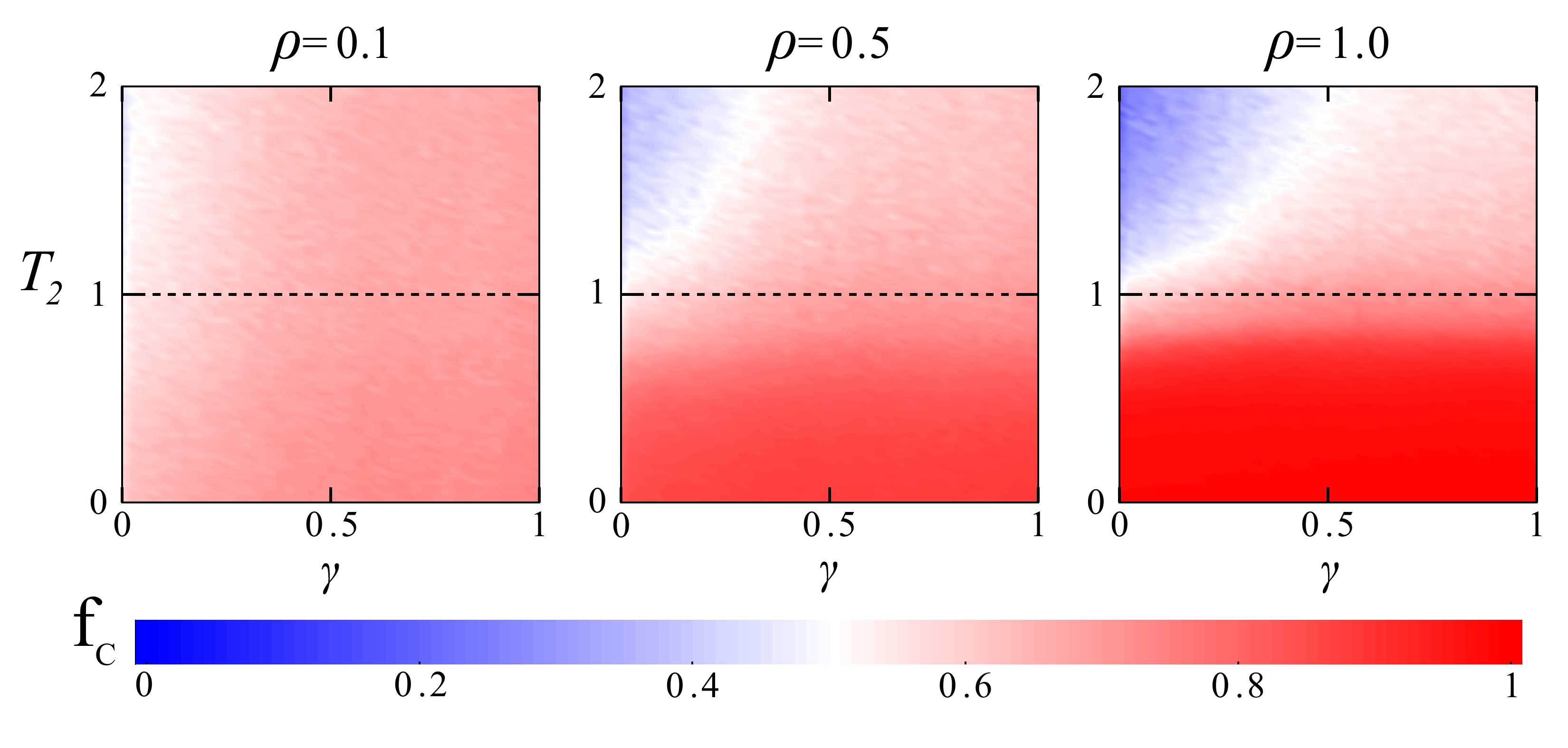}
        \caption{\textbf{Cooperation fraction diagrams.} Average cooperation fraction $f_C$ in $(\gamma,T_2)$ control parameter space across different higher-order interaction ratios: $\rho= 0.1$ (left), $0.5$ (center), and $1.0$ (right). In these panels, Game~1 and Game~3 are configured as Prisoner's Dilemma games, with parameters fixed at $T_1=T_3=1.2$ and $S_1=S_3=-0.2$. Game~2 transitions between a Prisoner's Dilemma, for $T_2\geq 1$, to a Stag Hunt game, when $T_2<1$, with fixed $S_2=-0.5$.}
    \label{fig:4}
\end{figure}

Figure~\ref{fig:4} illustrates the phase diagram of the average cooperation fraction $f_C$ in the control parameter space of $T_2$ and the reputation weight ($\gamma$) under varying proportions of higher-order interactions ($\rho=0.1$, $0.5$, $1.0$). The cooperation fraction is defined as the proportion of edges in the network where the strategy is cooperative ($s_{ij} = 0$), reflecting the prevalence of cooperative behavior in the system. Each point in these diagrams represents steady-state averages obtained from five independent network realizations.

Our findings reveal distinct cooperation regimes as Game~2 transitions from a Stag Hunt to a Prisoner's Dilemma. When Game~2 is an SH ($T_2 \in [0,1]$), the reputation weight $\gamma$ exerts limited influence on the fraction of cooperation. This insensitivity stems from the intrinsic stability of the cooperation-cooperation (CC) equilibrium in the SH, which naturally favors cooperative outcomes and diminishes the leverage of reputation feedback.

In contrast, when Game~2 is a PD ($T_2 \in [1,2]$), we observe the emergence of a defective majority—cooperation fractions below 0.5—particularly at low $\gamma$ values. The boundary of $f_C=0.5$ extends in a curvilinear fashion from $\gamma=0$ to $\gamma=0.5$ in the limit $\rho = 1.0$, delineating the transition between cooperation-dominated and defection-dominated states. This pattern highlights that direct reputation ($\gamma$ high) is substantially more effective than indirect reputation ($\gamma$ low) in sustaining cooperation. Direct reputation provides immediate behavioral feedback—rewarding cooperation and penalizing defection through real-time neighbor assessment. In contrast, indirect reputation relies on information transmitted through shared neighbors, which diffuses and aggregates across the network, making it less responsive to recent behavioral changes in dynamic environments.

The influence of higher-order interactions $\rho$ is strongly context-dependent and modulates these effects. In cooperative contexts (e.g., SH configurations), increasing $\rho$ amplifies cooperative signals and fosters the formation of cooperator clusters, further stabilizing the CC equilibrium. Conversely, in defection-prone settings (e.g., PD configurations), higher $\rho$ accelerates the spread of negative reputation, reinforcing defector dominance and precipitating a collapse of cooperation. These dynamics underscore the nonlinear coupling between structural complexity and reputation mechanisms: under favorable conditions, higher-order interactions amplify cooperation, whereas in competitive environments they magnify its breakdown.

\begin{figure}
\centering
\includegraphics[width=0.99\textwidth]{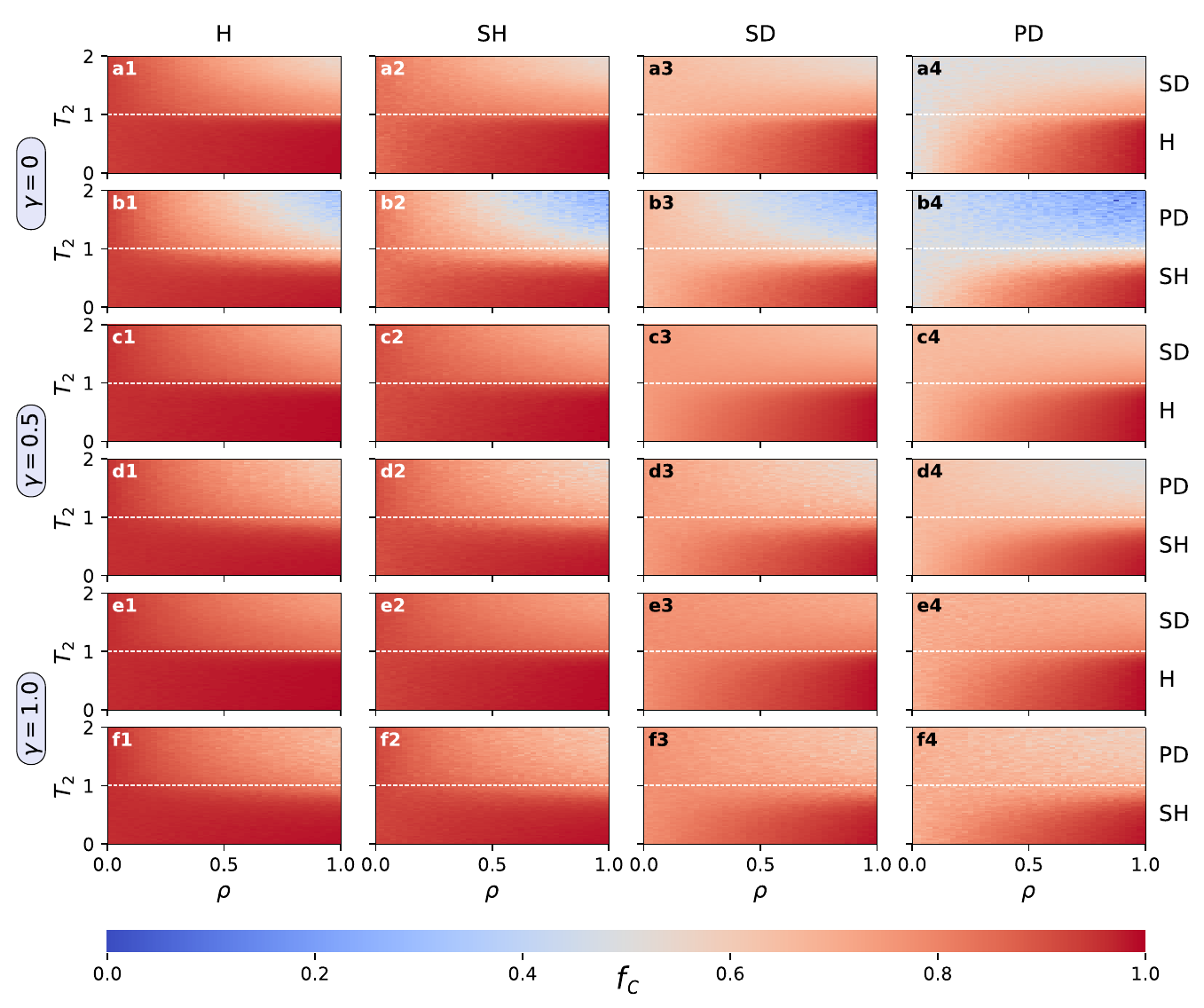}
\caption{\textbf{Cooperation fraction diagrams for varied game scenarios.} Contour plots of the cooperation fraction $f_C$ in the asymptotic state under the indirect reputation mechanism $\gamma=0$ (panels a and b), hybrid $\gamma=0.5$ (panels c and d), and direct $\gamma=1$ (panels e and f) are presented as a function of the $2$-simplex fraction $\rho$ and the payoff parameter $T_2$. Other parameters are specified as follows: the first column corresponds to the Harmony Game (H, with $T_1=T_3=0.8$, $S_1=S_3=0.2$); the second column corresponds to the Stag Hunt (SH, with $T_1=T_3=0.8$, $S_1=S_3=-0.2$); the third column corresponds to the Snowdrift Game (SD, with $T_1=T_3=1.2$, $S_1=S_3=0.2$); and the fourth column corresponds to the Prisoner's Dilemma (PD, with $T_1=T_3=1.2$, $S_1=S_3=-0.2$). Labels at the top of each column denote the respective types of Game~1 and Game~3. Additionally, each subfigure is divided into two regions by a horizontal dashed line at $T_2=1$, with labels on the right side indicating the corresponding type of Game~2 in each subpanel. For the upper panels, $S_2=0.5$, while for the lower panels, we set $S_2=-0.5$.}
\label{fig:5}
\end{figure}

Building upon the configuration analyzed in Fig.~\ref{fig:4}, where both Game~1 and Game~3 were fixed as Prisoner's Dilemmas and cooperation was examined across the $(\gamma, T_2)$ parameter space for selected values of $\rho$, we now extend the analysis to a broader set of base-game combinations and a complementary control space. In this experiment, we compute the macroscopic cooperation fraction $f_C$ in the $(\rho, T_2)$ plane for three reputation regimes-indirect ($\gamma=0$), hybrid ($\gamma=0.5$), and direct ($\gamma=1$)-to explore how the proportion of higher-order interactions ($\rho$) modulates cooperative dynamics under different informational pathways.

Figure~\ref{fig:5} presents this extended analysis in a multi-panel depiction. The columns, from left to right, correspond to Game~1 and Game~3 being configured as the Harmony Game (H), Stag Hunt (SH), Snowdrift Game (SD), and Prisoner’s Dilemma (PD), respectively. Panels~(a), (c), and~(e) adopt $S_2=0.5$, such that Game~2 represents a Harmony Game for $T_2 \in [0,1]$ and a Snowdrift Game for $T_2 \in [1,2]$, while Panels~(b), (d), and~(f) adopt $S_2=-0.5$, making Game~2 a Stag Hunt for $T_2 \in [0,1]$ and a Prisoner’s Dilemma for $T_2 \in [1,2]$. Compared to the absence of reputation mechanisms in \cite{guo2021evolutionary}, the inclusion of reputation—whether indirect, direct, or hybrid—profoundly reshapes the cooperative landscape. The figure reveals how informational pathways and structural coupling jointly determine the emergence and stability of cooperation across distinct game environments.

Under the indirect reputation mechanism $\gamma=0$ (panels a and b), cooperation levels are generally enhanced relative to the reputation-free baseline, but strong asymmetries emerge across game types. When Game~1 and Game~3 are both Harmony Games, low values of the higher-order interaction ratio ($\rho$) result in minimal influence from Game~2, and network strategies are mainly determined by the Nash equilibrium (CC) of Games~1 and~3. As $\rho$ increases, the influence of Game~2 becomes more pronounced. For $T_2\le1$, Game~2 corresponds to a Harmony ($S_2=0.5$) or Stag Hunt ($S_2=-0.5$) game, both favoring cooperation. Given the already high cooperation levels established by Games~1 and~3, increasing $\rho$ produces only marginal gains. Conversely, for $T_2\ge1$, Game~2 transitions to a Snowdrift ($S_2=0.5$) or Prisoner's Dilemma ($S_2=-0.5$), with equilibria biased toward defection ($CD/DC$ or $DD$). In these dynamic environments, the indirect reputation mechanism—based on information that propagates through intermediaries—responds sluggishly to behavioral shifts, and higher-order interactions amplify the spread of negative reputation, causing cooperation to decline as $\rho$ increases. Similar patterns appear across the SH, SD, and PD columns of Fig.~\ref{fig:5}, where variations in cooperation levels mirror the distinct Nash equilibria of each game type. Overall, indirect reputation promotes cooperation in stable settings dominated by equilibria favoring cooperation, but its reliance on diffused, multi-hop information limits its adaptability under competitive or rapidly changing conditions.

In contrast, the direct reputation mechanism $\gamma=1$ (panels e and f) exhibits a markedly stronger and more consistent capacity to sustain cooperation. As shown in Fig.~\ref{fig:5}, when $T_2<1$, direct reputation maintains a high level of cooperation across all $\rho$ values, regardless of the underlying game configuration of Games~1 and~3. Even when $T_2>1$, corresponding to the Snowdrift or Prisoner's Dilemma regimes, direct reputation effectively curbs the spread of defection, preventing the steep declines observed under indirect reputation. In configurations where Games~1 and~3 are Stag Hunt, Snowdrift, or Prisoner's Dilemma, the advantages of direct reputation become particularly evident: cooperation either remains stable or decreases only mildly with increasing $\rho$. These results confirm that the prompt responsiveness of direct reputation enhances system stability, enabling rapid detection and suppression of defection and thereby supporting robust cooperative clusters even in defection-prone environments.

The hybrid reputation mechanism, $\gamma=0.5$, (panels c and d) yields intermediate outcomes between the two extremes. By blending direct and indirect components, it partially compensates for the slower and more diffuse information processing of pure indirect reputation but does not match the stabilizing effects of fully direct feedback. When Games~1 and~3 are Harmony Games, the hybrid mechanism sustains cooperation levels between those observed under $\gamma=0$ and $\gamma=1$. In more competitive configurations (SH, SD, PD), the spatial patterns of $f_C$ under hybrid reputation closely resemble those of the indirect case, though with slightly weaker declines at high $\rho$. Overall, the hybrid mechanism can dampen the spread of defection but remains less effective than direct reputation in fostering widespread cooperation.

Taken together, the results of Fig.~\ref{fig:5} highlight a clear hierarchy of efficacy among reputation mechanisms. Indirect reputation enhances cooperation only in stable cooperative environments, hybrid reputation provides modest resilience by blending diffuse and immediate feedback, and direct reputation consistently promotes and stabilizes cooperation across all game contexts and structural regimes.

The analyses above demonstrate that cooperation emerges from a delicate balance between reputation mechanisms and higher-order structural effects. The nonlinear patterns observed in Fig.~\ref{fig:4} suggest that both positive and negative reputation feedbacks are amplified through higher-order interactions, depending on the underlying game configuration. To further clarify this coupling, we conduct a focused experiment by fixing the reputation weight at $\gamma=0.2$ and systematically examining how cooperative clusters form and evolve across varying $\rho$ and $T_2$ values.

\begin{figure}
\centering
\includegraphics[width=0.9\linewidth]{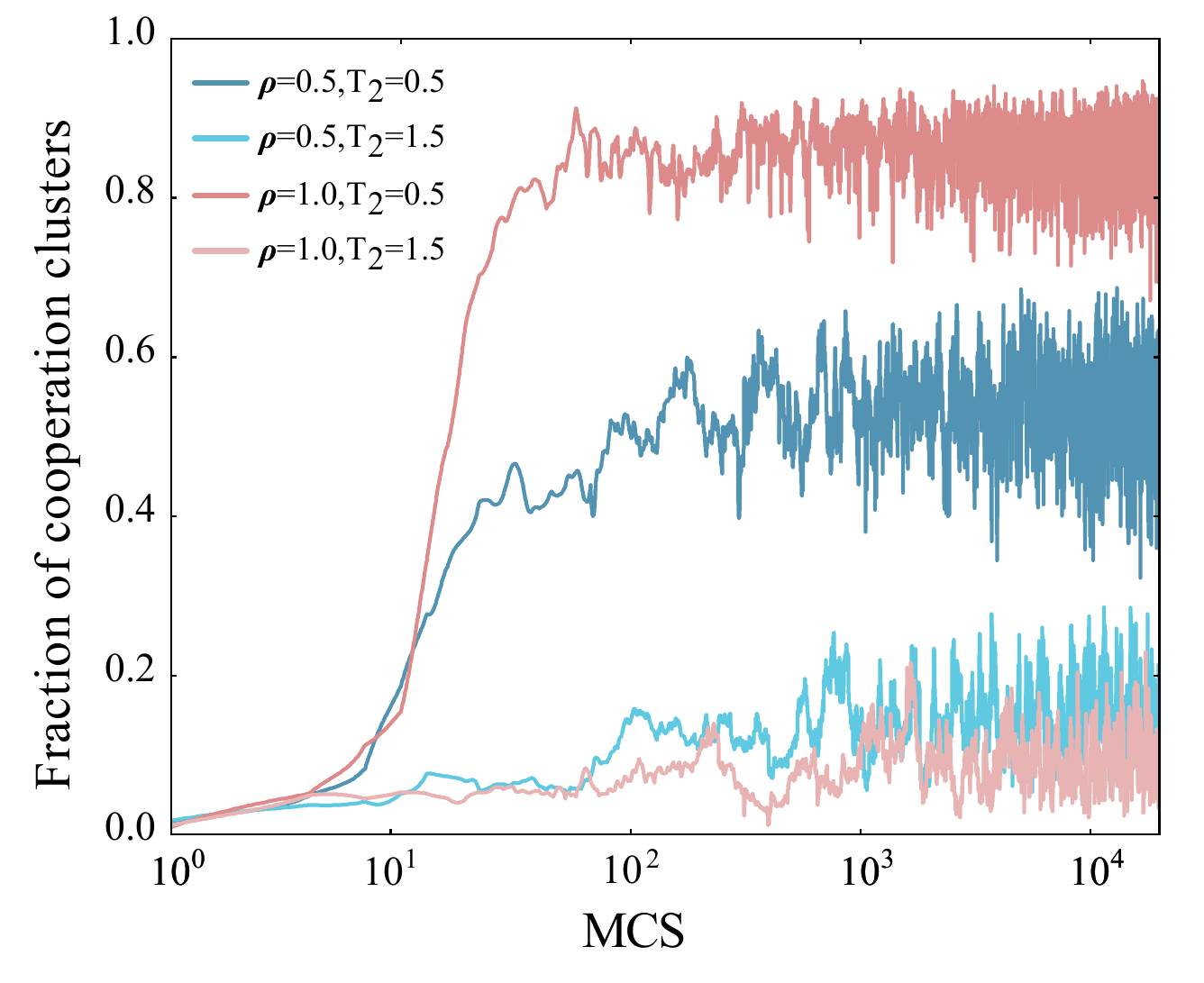}
\caption{\textbf{Temporal evolution of cooperator clustering.} The figure presents the time evolution of the proportion of cooperator clusters (defined as the ratio of fully cooperative triangles to total triangles) under four representative conditions: $\rho=0.5$ and $T_2=0.5$ (dark grey-blue), $\rho=0.5$ and $T_2=1.5$ (cyan), $\rho=1.0$ and $T_2=0.5$ (dark salmon), and $\rho=1.0$ and $T_2=1.5$ (light salmon).}
\label{fig:6}
\end{figure}

In this analysis, each triangular structure (denoted as $\Delta\text{Tr}^{\text{one}}$ and $\Delta\text{Tr}^{\text{two}}$) represents a group of three interacting individuals, with the triangle's vertices corresponding to individuals $i$, $j$, and $k$. A group is identified as a cooperator cluster when all three members adopt cooperative strategies toward one another. Figure~\ref{fig:6} shows the temporal evolution of the proportion of such cooperator clusters under different combinations of $\rho$ and $T_2$.

When $T_2=0.5$, representing a low temptation to defect, the fraction of cooperator clusters rises sharply with increasing $\rho$, indicating that higher-order interactions reinforce cooperative reciprocity. In contrast, when $T_2=1.5$, corresponding to a high temptation to defect, the fraction of cooperator clusters decreases as $\rho$ grows. This reversal illustrates the dual role of higher-order structures: in cooperative contexts, they amplify positive reputation and facilitate cohesive cooperator clusters; in defection-prone contexts, they intensify the transmission of negative reputation, accelerating the erosion of cooperation.

These findings provide a microstructural confirmation of the macroscopic patterns previously observed, thus demonstrating that the nonlinear coupling between reputation and higher-order interactions not only shapes aggregate cooperation levels but also governs the spatial organization of cooperative behavior. This multi-scale interplay offers new theoretical insight into how reputation dynamics and network reciprocity jointly drive the evolution of cooperation in complex social systems.

\section{Conclusion}
\label{sec:conclusion}

Although cooperative behavior seems to conflict with self-interested tendencies shaped by natural selection, five mechanisms provide theoretical foundations for its evolution. In particular, network reciprocity shows how specific topologies can enhance cooperation~\cite{nowak1992evolutionary}. To deepen understanding of cooperation's emergence, recent work examines how reputation mechanisms interact with network structure.

In this study, we proposed an edge-based, vector-valued reputation mechanism and integrated it with higher-order network structures to investigate how reputation influences the collective evolution of strategies. Using evolutionary game theory, we analyzed environments with both pairwise and triadic interactions, where nodes select role models for strategy learning based on reputation, thereby affecting strategy updates and facilitating the emergence of cooperation.

Our analysis yields several insights. First, by comparing direct, indirect, and hybrid reputation mechanisms across social dilemmas with different Nash equilibria, we find that direct reputation most effectively promotes cooperation in dynamic settings. Indirect reputation, which aggregates information from common neighbors, enhances cooperation in more stable environments but is limited by its slower, network-mediated information propagation in dynamic ones. The hybrid mechanism, combining both components, performs between the two extremes, suggesting that mixed feedback capturing both local and global information is a useful approximation for social systems.

Second, examining how the structural parameter $\rho$ regulates cooperator clustering under varying dilemma intensities reveals a nonlinear coupling between higher-order structure and reputation. When the temptation to defect is low, higher-order interactions amplify cooperative signals and strengthen cooperator clustering. When the temptation is high, the same structural coupling accelerates the diffusion of negative reputation, eroding cooperation. These findings underscore that the joint action of topology and reputation governs not only macroscopic cooperation levels but also the microstructural organization of cooperative clusters.

Despite these advances, several limitations remain. The present model assumes a static higher-order topology, whereas real social and organizational networks continuously evolve as agents form and dissolve interactions. Future research should therefore explore the co-evolution of reputation and higher-order connectivity, allowing both to adapt dynamically. Moreover, our formulation treats transmitted reputation as accurate information, yet in realistic settings, reputation signals may be noisy, delayed, or strategically distorted. Introducing stochastic or biased information exchange could reveal how misinformation or perceptual errors reshape cooperative equilibria. Finally, extending this framework to heterogeneous or multiplex settings—where different layers of interaction (e.g., social, economic, and informational) coexist—would further strengthen its explanatory and predictive power. Such directions would provide a more comprehensive account of how cooperative behavior emerges, stabilizes, and transforms within complex social systems.

\section*{Acknowledgments}

C. P. is funded by the Yunnan Fundamental Research Projects (No.202401AU070018), the Scientific Research Fund of the Yunnan Provincial Department of Education (No. 2024J0774). 
Y.K. acknowledges support from the Yunnan Fundamental Research Projects (Grant Nos. 202501AU070193), and the Scientific Research Fund of the Yunnan Provincial Department of Education (Grant No. 2025J0579). Y. M. was partially supported by the Government of Aragon, Spain, and ERDF "A way of making Europe" through grant E36-23R (FENOL), and by Grant No. PID2023-149409NB-I00 from Ministerio de Ciencia, Innovación y Universidades, Agencia Española de Investigación (MICIU/AEI/10.13039/501100011033) and ERDF ``A way of making Europe''.

\bibliographystyle{elsarticle-num} 
\bibliography{references}

\end{document}